\documentclass[12pt,a4paper]{article}
\usepackage{epsfig}
\renewcommand{\thefootnote}{\fnsymbol{footnote}}
 
\def\eq#1{{Eq.~(\ref{#1})}}

\newcommand{\be}{\begin{equation}}
\newcommand{\ee}{\end{equation}}
\newcommand{\beqar}[1]{\begin{eqnarray}\label{#1}}
\newcommand{\eeqar}{\end{eqnarray}}
\newcommand{\ben}{\begin{eqnarray*}}
\newcommand{\een}{\end{eqnarray*}}

\newcommand{\as}{\alpha_s}
\newcommand{\un}{\underline}

\newcommand{\stackeven}[2]{{{}_{\displaystyle{#1}}\atop\displaystyle{#2}}}
\newcommand{\lsim}{\stackeven{<}{\sim}}
\newcommand{\gsim}{\stackeven{>}{\sim}}
\begin{document}
\title {{\bf Thoughts on Non-Perturbative Thermalization\vspace*{7mm} 
and Jet Quenching in Heavy Ion Collisions
\\[1cm] }}
\author{{\bf Yuri V. Kovchegov\thanks{e-mail: yuri@mps.ohio-state.edu}}~\\[10mm]
{\it\small Department of Physics, The Ohio State University}\\
{\it\small Columbus, OH 43210} }
 
\date{July, 2005}
\maketitle
\thispagestyle{empty}
 
\begin{abstract}
We start by presenting physical arguments for the impossibility of
perturbative thermalization leading to (non-viscous) Bjorken
hydrodynamic description of heavy ion collisions. These arguments are
complimentary to our more formal argument presented in
\cite{hydro}. We argue that the success of hydrodynamic models in 
describing the quark-gluon system produced in heavy ion collisions
could only be due to non-perturbative strong coupling effects. We
continue by studying non-perturbative effects in heavy ion collisions
at high energies. We model non-perturbative phenomena by an instanton
ensemble. We show that non-perturbative instanton vacuum fields may
significantly contribute to jet quenching in nuclear collisions.  At
the same time, the instanton ensemble contribution to thermalization
is likely to be rather weak, leading to non-perturbative
thermalization time comparable to the time of hadronization. This
example illustrates that jet quenching is not necessarily a signal of
a thermalized medium. Indeed, since the instanton models do not
capture all the effects of QCD vacuum (e.g. they do not account for
confinement), there may be other non-perturbative effects facilitating
thermalization of the system.
\end{abstract}
\thispagestyle{empty}

\newpage

\renewcommand{\thefootnote}{\arabic{footnote}}
\setcounter{footnote}{0}

\setcounter{page}{1}
%

\section{Introduction}

The problem of thermalization has become one of the central questions
in the theory of heavy ion collisions. The success of hydrodynamic
models \cite{bj,EKR,hydro1,hydro2,HN} in describing RHIC data on
elliptic flow \cite{flow}, along with the suppression of high-$p_T$
particles in $Au+Au$ collisions as compared to $d+Au$ and $pp$
collisions
\cite{dAtaphen,dAtaphob,dAtastar,brahms,aaphenix,aaphobos,aastar,Bj,EL,BDMPSfull,EL2,Zak,SW},
indicate creation of a strongly interacting and likely thermal medium
in the final state of heavy ion collisions --- the quark-gluon plasma
(QGP). Currently, the main theoretical puzzle is in understanding how
this medium was formed in the collision.

The initial conditions for creation of QGP has received a lot of
attention in the recent years \cite{KMW,KR,GM,KV,K2,KNV,Lappi} in the
framework of saturation/Color Glass Condensate (CGC) physics
\cite{glr,mq,bm,mv,k1,jkmw,dip,bk,jimwlk,GM1,JMK}. The data on hadron 
multiplicity and spectra obtained in $d+Au$ collisions at RHIC
\cite{brahms-1,brahms-2,phenix,phobos,star} appears to be in good
agreement with expectations of Color Glass approach
\cite{KLNd,klm,kkt1,aaksw,bkw,jmnv,km,dmc,kt,braun}. When
combined with the success of saturation models \cite{KN} in describing
total charged hadron multiplicity in $Au+Au$ collisions at RHIC, this
allows one to hope that evolution of the produced quark-gluon system
toward thermalization could also be described within the
saturation/Color Glass framework, which indeed involves using Feynman
diagrams.

However, understanding perturbative thermalization proved to be a very
hard problem. Several perturbative thermalization scenarios have been
proposed in the recent years
\cite{BMSS,Mueller_eq,SS,BV,DG,MG,Wongc}. None of them appears to lead
to thermalization time as short as required by RHIC phenomenology,
which is $\tau \lsim 0.5$~fm/c. Moreover, the so-called "bottom-up"
thermalization scenario, which was originally proposed in \cite{BMSS}
and was the first one to use saturation initial conditions, proved to
be susceptible to plasma instabilities
\cite{Uli,Mrow,ALM,AL,RM,SR,MMR,ALMY}. This brought hope that
instability effects would speed up the onset of
thermalization. However, in \cite{AL} it was shown that instabilities
(along with any other dynamical effects) obey a lower bound on
thermalization time, which is parametrically very close to (though
slightly lower than) the thermalization time obtained in the
"bottom-up" scenario \cite{BMSS}. More recent developments include
suggestion to consider the onset of isotropization instead of
thermalization as the necessary condition for hydrodynamic models to
work \cite{ALMY}, a numerical analysis of instabilities demonstrating
that their magnitudes saturate as time progresses \cite{AMY2}, leading
to, probably, a less severe impact on the system as thought
previously, and a new modified version of the "bottom-up" scenario
\cite{MSW}.

In our recent paper on the subject \cite{hydro} we considered the
onset of thermalization as a possible transition from the $\epsilon
\sim 1/\tau$ scaling of the energy density characteristic of the Color
Glass condensate initial conditions \cite{KMW,KR,GM,KV,K2,KNV,Lappi},
to a steeper falloff with proper time, $\epsilon \sim
1/\tau^{1+\Delta}$, expected if the system obeys ideal Bjorken
hydrodynamics \cite{bj} ($\Delta > 0$). We were looking for
perturbative corrections, which could be, for instance, powers of
$\Delta \, \ln \tau$, with $\Delta$ being proportional to some power
of the strong coupling constant $\as$. Having found no such (or any
other) important corrections to $1/\tau$ scaling of energy density we
proved that no corrections of this type are possible in perturbation
theory. Therefore, we proved that perturbative thermalization is
impossible in heavy ion collisions. Our arguments in \cite{hydro} were
somewhat involved, and lacked a clear physical argument demonstrating
impossibility of perturbative thermalization. Below we try to
construct such an argument in Section 2. There we provide two
arguments. The first one, presented in Sect. 2.1, demonstrates that,
in the absence of non-perturbative effects, if the theory is defined
only perturbatively, Bjorken hydrodynamics would not lead to
decoupling of the system. Therefore, plane-wave final states necessary
in perturbation theory to define scattering amplitudes and particle
production cross sections are never achieved. This indicates a
contradiction between Bjorken hydrodynamics and perturbation theory,
indicating possible impossibility of obtaining the former from the
latter. The discrepancy becomes apparent in the second argument,
presented in Sect. 2.2. There we argue that, in the zero coupling
limit ($g \rightarrow 0$) taken at some late time after the collision,
QCD perturbation theory would indeed lead to "free-streaming" with the
energy density scaling like $\epsilon \sim 1/\tau$, while Bjorken
hydrodynamics would give $\epsilon \sim 1/\tau^{4/3}$. The formal
reason for such scaling in the case of Bjorken hydrodynamics is that
the (non-interacting) ideal gas equation of state is $\epsilon = 3 p$,
and this is all that's needed to obtain the $\epsilon \sim
1/\tau^{4/3}$ scaling. The physical reason for $\epsilon \sim
1/\tau^{4/3}$ scaling in the $g \rightarrow 0$ limit is that, in
writing $\epsilon = 3 p$ equation of state for the ideal gas, one
indeed assumes a presence of some thermal bath (e.g. the box
containing the particles) with which the gas particles interact and
which keeps them thermal throughout the evolution of the system. Such
external thermal bath is absent in heavy ion collisions. Therefore,
Bjorken hydrodynamics is impossible to achieve in the weak coupling
limit. We further argue that another way to reach this conclusion is
by noticing that shear viscosity, which is the first non-equilibrium
correction to Bjorken hydrodynamics, diverges in the weak coupling
limit (see \eq{vis}) \cite{AMY}, indicating that non-equilibrium
corrections become comparable to the equilibrium values of the
pressure, breaking down the ideal (non-viscous) hydrodynamics in the
weak coupling limit. Our conclusion here is that thermalization of the
quark-gluon system (if it does take place) leading to hydrodynamic
description of the heavy ion collisions could only be due to
non-perturbative strong coupling effects.

Indeed, understanding the non-perturbative phenomena in the
non-equilibrium setting of heavy ion collisions is a very difficult
(if not presently impossible) task. Here we will consider one
well-studied model of non-perturbative effects --- the instanton
vacuum model \cite{BPST,valley,DP,RS,size,TE,CDG}. Since, as we argue
in Sect. 2, non-perturbative effects appear to be important in $AA$
collisions for thermalization, which is required by phenomenology to
describe the observed data on elliptic flow \cite{hydro1,hydro2,flow},
we would like to investigate their influence on other important
observables in heavy ion collisions. In Sect. 3 we study the effect of
the instanton vacuum on jet quenching and thermalization in heavy ion
collisions. Reviewing the results of \cite{KKL} in Sect. 3.1 we argue
that the instanton density is suppressed by strong perturbative gluon
fields produced (in the Color Glass Condensate framework) in the early
stages of the collision with the proper times $\tau \lsim 1/Q_s$. At
slightly later proper times, for $\tau \gsim 1/Q_s$, perturbative
fields linearize and instanton vacuum is not suppressed anymore. At
this point we would have a system of free quarks and gluons
propagating through the instanton vacuum. Usually, a parton produced
in a collision and propagating as a free plane wave would start
hadronizing right away. This is probably true in $pp$ or $e^+ e^-$
collisions. However, in $AA$ collisions, at times $\tau
\gsim 1/Q_s$, the energy density is still too high for hadronization
to take place. Unable to hadronize due to high energy density,
non-interacting quarks and gluons are likely to interact with the
instantons in the vacuum and to loose energy as they propagate through
the instanton ensemble until the hadronization time. (Once quarks and
gluons are converted into hadrons they, of course, do not loose energy
in the vacuum anymore.) The mechanism of parton-instanton interactions
has been extensively studied in \cite{heinst}. One might expect the
energy loss of partons in the instanton vacuum to be negligibly small
due to low density of instantons, which would play the role of
scattering centers. However, the peculiar property of the energy loss
formula derived in \cite{BDMS,GVWZ} and shown here in \eq{de} is that
the integrated partonic energy loss is particularly sensitive to late
proper times of the collision due to an extra factor of $\tau$ in the
integrand. Namely, it is not as important what the energy density was
in the early stages of the collision, as it matters what the energy
density is at late proper times. Therefore, the energy loss of a fast
moving parton in a non-static medium is mostly determined by the
medium density at late times. Since, at late proper times close to the
time of hadronization, the density of scattering sources in the
produced medium is comparable to the density of instantons in the
vacuum, the latter may significantly contribute to jet quenching, on
par with traditional energy loss in the produced medium, as we
demonstrate by parametric estimates in Sect. 3.2.

At the same time, as the vacuum instanton density is low, its
contribution to thermalization appears to be rather weak, leading to
non-perturbative thermalization time comparable to the time of
hadronization, as we show in Sect. 3.3. This example illustrates that
jet quenching is not necessarily a signal of a thermalized medium and,
if \eq{de} adequately describes energy loss in a heavy ion collision,
one may have significant jet quenching without thermalization. Of
course, as we argue in Sect. 4, since the instanton models do not
capture all the effects of QCD vacuum (e.g. they can not explain
confinement), there may be other non-perturbative effects, such as QCD
strings \cite{strings}, facilitating thermalization of the system and
leading to energy loss in the non-perturbatively formed QGP.

\section{Non-perturbative Nature of Bjorken Hydrodynamics}

In this Section we will present two qualitative arguments showing that
perturbative thermalization leading to Bjorken hydrodynamics
description of heavy ion collisions is impossible.

\subsection{Argument I}

Before we begin, let us briefly review the main assumptions of Bjorken
hydrodynamics \cite{bj}. The main idea behind the ideal (non-viscous)
Bjorken hydrodynamics is that the medium produced in a collision of
two very large nuclei can be described by hydrodynamic equations
\be\label{tcons}
\partial_\mu \, T^{\mu\nu} \, = \, 0
\ee
with the energy-momentum tensor given by
\be\label{hydro}
T_{\mu\nu} \, = \, (\epsilon + p) \, u_\mu \, u_\nu - p \, g_{\mu\nu},
\ee
where, in the boost-invariant (rapidity-independent) case the fluid
velocity is
\be
u_\pm \, = \, \frac{x_\pm}{\tau}, \hspace*{1cm} {\un u} \, = \, 0
\ee
and the proper time is $\tau = \sqrt{2 x_+ x_-}$. We use the light
cone coordinates
\be
x_\pm = \frac{t \pm z}{\sqrt{2}}
\ee
and $\un u$ denotes a two-dimensional transverse vector. $\epsilon$ is
the energy density and $p$ is the pressure of the fluid. Substituting
\eq{hydro} into \eq{tcons} and assuming that $\epsilon = \epsilon (\tau)$ 
and $p = p (\tau)$ yields \cite{bj}
\be\label{hydroeq}
\frac{d \epsilon}{d \tau} \, = \, - \frac{\epsilon + p}{\tau}.
\ee
\eq{hydroeq} leads to the scaling of energy density with the proper time \cite{bj}
\be\label{etau}
\epsilon \, \sim \, \frac{1}{\tau^{1+\Delta}}
\ee
where
\be\label{dconst}
0 < \Delta \le \frac{1}{3}.
\ee
The value of $\Delta = 1/3$ can be obtained in the ideal
(ultrarelativistic) gas case with the equation of state
\be\label{e3p}
\epsilon \, = \, 3 \, p
\ee
and leads to 
\be\label{etauid}
\epsilon \, \sim \, \frac{1}{\tau^{4/3}}.
\ee
The value of $\Delta =0$, leading to 
\be
\epsilon \, \sim \, \frac{1}{\tau}
\ee
could in principle be achieved in a QCD plasma near the critical phase
transition temperature $T_c$ if the phase transition was first order
\cite{bj,KL}. However, the confining phase transition in QCD is non-perturbative: 
therefore it is impossible to obtain QCD matter at $T = T_c$ through
Feynman diagrams and we will not be concerned with this case here.

Now, let us assume that, as a result of summation of perturbative
diagrams to any order (and to all orders if necessary) one shows that
in a heavy ion collision a thermalized medium is produced which is
describable by Bjorken hydrodynamics. With logarithmic (due to running
of the coupling) accuracy the energy density of the resulting medium
would scale with the temperature $T$ as
\be\label{et4}
\epsilon \, \sim \, T^4.
\ee
This formula is better justified in the weak coupling limit, where the
diagrammatic analysis is at its best too. Comparing \eq{et4} to
\eq{etau} yields
\be\label{ttau}
T \, \sim \, \frac{1}{\tau^\frac{1 + \Delta}{4}}.
\ee
For example, for ideal gas \eq{ttau} yields $T \, \sim \,
\tau^{-1/3}$. 

Let us now imagine that thermalization and the subsequent evolution of
the system are completely described by the perturbation theory. In
other words, let us imagine ``turning off'' confinement and all other
effects associated with the non-perturbative QCD scale $\Lambda_{QCD}$
and assume that the theory is defined perturbatively. Indeed this is
not a realistic assumption, but would help us clarify whether
thermalization in heavy ion collisions is of perturbative nature or
not: if thermalization is perturbative we should not get any
inconsistencies by ``turning off'' confinement effects. If we do get
inconsistencies, then perturbative thermalization would be impossible,
and if thermalization does take place in heavy ion collisions, it
could only be due to non-perturbative effects.

For simplicity let us consider only gluons and massless quarks in the
system. Since the temperature $T$ is now the only scale in the
problem, the perturbative mean free path of a (massless) quark or a
gluon in the medium would scale like
\be\label{lt}
\lambda_{pert} \, \sim \, \frac{1}{T},
\ee
which, using \eq{ttau} gives
\be\label{ltau}
\lambda_{pert} \, \sim \, \tau^\frac{1 + \Delta}{4}.
\ee
Employing \eq{dconst} we obtain an upper limit on the mean free path
\be\label{lconst}
\lambda_{pert} \, \sim \, \tau^\frac{1 + \Delta}{4} \, \le \, \tau^{1/3}.
\ee
The equality in \eq{lconst} is achieved only for an ideal gas with the
equation of state given by (\ref{e3p}). (In \eq{lconst} we have
omitted $\tau$-independent factors which would make the inequality
dimensionally correct: putting them back into the formula would not
modify the conclusion, but would unnecessarily complicate the
equation. The same applies to \eq{nodec} below.)

Noting that the typical longitudinal extent of the system at time $t$
after a heavy ion collision is
\be
L \, = \, t \, \ge \, \tau
\ee
we conclude that 
\be\label{nodec}
\lambda_{pert} \, \le \, \tau^{1/3} \, < \, \tau \, \le \, L. 
\ee
(The transverse extent of the system is assumed to be infinitely large
\cite{bj} and is, therefore, always larger than the mean free path.) 

\eq{nodec} implies that the mean free path of a gluon or a massless 
quark in a plasma described by Bjorken hydrodynamics is always smaller
than the size of the system. Therefore, the decoupling condition
\be
\lambda \, \sim \, L
\ee
is {\sl never achieved} in perturbative Bjorken plasma.

(Indeed, once we turn the confinement effects back on, we would get
another scale in the problem, $\Lambda_{QCD}$, which would lead to
\be\label{nplam}
\lambda_{non-pert} \, = \, \frac{1}{\sigma \, \rho} \, \sim \, 
\frac{1}{\frac{1}{\Lambda^2_{QCD}} \, T^3} \, \sim \, \tau^{\frac{3 (1+\Delta)}{4}}
\ee
giving the mean free path 
\be
\lambda_{non-pert} \, \sim \, \tau \, \sim \, L
\ee
for the ideal gas case ($\Delta = 1/3$), which marginally satisfies
the decoupling condition. Here $\sigma$ is the parton-parton
scattering cross section and $\rho$ is the density of matter.)

As we can see from \eq{nodec}, the asymptotic $\tau \rightarrow +
\infty$ state of a perturbative plasma described by Bjorken hydrodynamics 
is not ``free streaming'': in fact, the inequality in \eq{nodec} gets
more justified as $\tau$ increases. Each individual parton can escape
the system with an exponentially small probability, $e^{-L/\lambda}
\sim \exp ( - \mbox{const} \, \tau^{2/3})$, which only decreases with time. 
This behavior is in contradiction with the diagrammatic approach: the
final state of any Feynman diagram in high energy scattering is a set
of free non-interacting particles flying off to infinity. The
existence of such a plane wave final state is an essential assumption
of the perturbation theory. Without the plane wave final state the
problem of particle production would be ill-defined. If diagrams would
lead to Bjorken hydrodynamics, than, such a common observable as the
one-gluon inclusive production cross section would be impossible to
define and calculate in $AA$ collisions, since there will be no
plane-wave particles in the final state. Single gluon inclusive
production cross section is well-defined in the quasi-classical
approximation of McLerran-Venugopalan model \cite{mv,KMW,KR,GM,KV,K2},
which does not led to a thermalized final state \cite{KV}. If
thermalization comes in through the higher order quantum corrections
to classical fields \cite{BMSS}, such corrections would invalidate the
notion of one-gluon inclusive production cross section, which appears
absurd.  Therefore, perturbative diagram resummation {\sl can not}
lead to Bjorken hydrodynamics: diagrams lead to a free streaming final
state, while Bjorken hydrodynamics in the absence of confinement is an
eternally interacting state.

Of course, for realistic finite nuclei, when the distance between the
nuclei becomes comparable to their radii, the expansion becomes
three-dimensional: in this sense Bjorken hydrodynamics is not really
an ``eternally'' interacting state.  The onset of 3D expansion would
indicate a transition from Bjorken to Landau hydrodynamics and is
outside of the scope of this discussion. It would be very unlikely
though for thermalization to happen at this late stage of the
collision when the system becomes very dilute and the energy density
starts falling much faster with time than it was in the Bjorken
hydrodynamics stage of the collision.

One may suggest that Feynman diagrams may lead to Bjorken
hydrodynamics in some intermediate state preceding free
streaming. Again this is not possible, since, as was shown above,
Bjorken hydrodynamics does not lead to free streaming, and if the
system enters a Bjorken hydrodynamics state it would never leave
it. Therefore Bjorken hydrodynamics can not be an intermediate state
of the system.

Introducing massive quarks in the argument is not going to change the
conclusions: if the quark mass is much smaller than the temperature,
$m_q \ll T$, then it could be neglected yielding $\sigma \sim \as^2 /
\as T^2 \sim 1/T^2$ and $\rho \sim T^3$ leading to the mean free path of
\eq{lt}. If quark mass is much bigger than the temperature, $m_q \gg T$, then 
the particle density would still be dominated by gluons yielding $\rho
\sim T^3$ and the interaction cross sections would also be dominated by
gluon exchanges, giving $\sigma \sim 1/T^2$, and we would recover
\eq{lt} again.

The argument presented above shows that Bjorken hydrodynamics is
fundamentally different, for instance, from the case of Landau
hydrodynamics. In the latter case decoupling is possible even in the
perturbative case. For simplicity let us consider an expanding
three-dimensional spherical droplet with the energy density within the
droplet independent of spacial coordinates and with the ideal gas
equation of state (\ref{e3p}). Then for such a system
\be
\epsilon \, \sim \, \frac{1}{t^4}. 
\ee
Following the above steps we obtain, using \eq{et4}, that $T \sim 1/t$
and the mean free path is
\be
\lambda_{pert} \, \sim \, \frac{1}{T} \, \sim \, t.
\ee
Since the radius of an ultrarelativistic droplet is $R \sim t$ we see
that $\lambda_{pert} \, \sim \, R$ and the system may decouple at late
times. Therefore, thermalization in this system, leading to Landau
hydrodynamics description of its evolution, appears to be possible to
describe using Feynman diagrams as an intermediate state between the
creation of the system and eventual decoupling. Indeed to make a
definitive conclusion on the subject of perturbative thermalization in
3D expanding system one needs to perform a careful diagrammatic
analysis similar to the one carried out in \cite{hydro} for the case
of Bjorken hydrodynamics.

One may worry that, for instance, Boltzmann equation has a
diagrammatic interpretation \cite{MS}, i.e., it does sum up certain
classes of diagrams in its collision term, and, nevertheless, it does
lead to perturbative thermalization. However we would like to point
out that Boltzmann equation leads to thermalization for a system of
finite volume, such as a gas of particles in a box. If a system is not
confined to a fixed volume, the question of thermalization of the
system in the weak coupling limit due to Boltzmann equation dynamics
is still an open question which has to be addressed on individual
basis for each system.

For the case of a system in a box, thermalization time in the weak
coupling limit is comparable to the size of the box. In the
diagrammatic language this implies that if we iterate the diagrams
from the collision term of the Boltzmann equation to obtain a full
resummed diagram describing the system, such diagram, covering the
evolution of the system over a very long time scale in a fixed finite
volume, should also include information on interactions of particles
with the box in which they are confined, and, because of that, would
not have an asymptotic ``free streaming'' final state. Therefore, it
would not be a diagram in the vacuum and the above argument would not
apply to it: one can think of the box as of an external field, which
is present at all times preventing decoupling of particle interactions
and not allowing them to free stream to spacial infinity.

To conclude this Section let us reiterate our argument once
more. Bjorken hydrodynamics \cite{bj}, in the weak coupling limit (in
the absence of confinement effects), does not decouple: gluons and
quarks continue interacting indefinitely, as long as one-dimensional
expansion continues. Such system does not have a free final state,
and, therefore can not be obtained from Feynman diagrams in vacuum,
which, at any finite order in the coupling, and, therefore, to all
orders as well, always do have a non-interacting final state. (Indeed
Feynman diagrams in an external field, such as the graphs describing a
system of particles in a box, do not have an asymptotically free final
state: however, since, of course, there is no external field in heavy
ion collisions, such diagrams would be irrelevant here.)

\subsection{Argument II}

Now let us assume that the system of quarks and gluons produced in a
heavy ion collision reaches equilibration at some proper time
$\tau_{th}$. Let us pick some proper time $\tau_0$ after
thermalization, $\tau_0 \ge \tau_{th}$, and set the QCD coupling
constant $g$ to be equal to zero for all times after $\tau_0$, i.e.,
$g=0$ for $\tau \ge \tau_0$. Below we will compare the $g \rightarrow
0$ limits of the full theory (QCD) and of Bjorken hydrodynamics.

First of all, from QCD standpoint, if the coupling goes to zero, $g
\rightarrow 0$, the system would stop interacting and quarks and gluons 
would ``free-stream'' away. The energy density of a free-streaming
quarks and gluons scales as
\be\label{e1t}
\epsilon \, \sim \, \frac{1}{\tau}
\ee
simply due to energy conservation in a one-dimensionally expanding
system. (To obtain a different energy scaling with proper time, such
as in Bjorken hydrodynamics case in \eq{etau}, the system needs to
perform work, which obviously does not happen with a free-streaming
system.)

To analyze the $g \rightarrow 0$ limit of Bjorken hydrodynamics we
first note that it requires only conservation of the energy-momentum
tensor (Eqs. (\ref{tcons}) and (\ref{hydro})) with an equation of
state relating $\epsilon$ and $p$. Since, in QCD, perturbative
equation of state is known and is given by \cite{EOS}
\be
\frac{\epsilon}{3 \, p} \, = \, 1 + o (g^2),
\ee
we naturally conclude that in $g \rightarrow 0$ limit it reduces to
the ideal gas equation of state (\ref{e3p}). Using that equation of
state in \eq{hydroeq} governing Bjorken hydrodynamics \cite{bj} leads
to
\be\label{ebt}
\epsilon \sim \frac{1}{\tau^{4/3}}.
\ee
Thus we have established that the $g \rightarrow 0$ limits of a field
theory, such as QCD, and of Bjorken hydrodynamics are {\sl different},
given by Eqs. (\ref{e1t}) and (\ref{ebt}) correspondingly. Therefore,
perturbative QCD can never lead to ideal Bjorken hydrodynamics. 

The origin of the difference between the two limits is clear: the
ideal gas equation of state (\ref{e3p}) describes a gas of particles
which do not interact with each other, but which interact with some
thermal bath. This thermal bath could be a box the particles are
confined to or some external field. Indeed there is no external
thermal bath in the case of a heavy ion collision: hence Bjorken
hydrodynamic approximation has a different $g \rightarrow 0$ limit
from the full theory.

One may worry that putting the coupling constant to zero at proper
time $\tau \ge \tau_0$ would violate gauge invariance. Indeed this is
true in general making our argument not quite rigorous. However, as
was argued convincingly in the existing thermalization scenarios
\cite{BMSS,Mueller_eq,SS,BV,DG,MG,Wongc}, by the time of possible
thermalization (and even much earlier) the density of particles would
become sufficiently low for the quarks and gluons to go on mass shell
between the interactions. This would insure gauge invariance of the
quark and gluon distributions at any given late proper time.

To reconcile the $g \rightarrow 0$ limit of Bjorken hydrodynamics from
\eq{ebt} with the more physically intuitive scaling of \eq{e1t} one has 
to include {\sl non-equilibrium} viscous corrections
\cite{DGy,Teaney}. Introducing shear viscosity dependence into
the ideal Bjorken hydrodynamics energy-momentum tensor would make it
depend on three variables --- $\epsilon$, $p$ and $\eta$, and is
accomplished by adding a viscosity-dependent term to \eq{hydro}. The
modified energy-momentum tensor for a baryon-free fluid is
\cite{DGy,AMY,Teaney}
\be\label{hydrovis1}
T_{\mu\nu} \, = \, (\epsilon + p) \, u_\mu \, u_\nu - p \, g_{\mu\nu}
+ \eta \, \left( \nabla_\mu u_\nu + \nabla_\nu u_\mu - \frac{2}{3} \,
\Delta_{\mu\nu} \, \nabla_\rho \, u^\rho \right),
\ee
where $\nabla_\mu \equiv (g_{\mu\nu} - u_\mu \, u_\nu) \,
\partial^\nu$ and $\Delta_{\mu\nu} \equiv g_{\mu\nu} - u_\mu \, u_\nu$. 
The shear viscosity in perturbation theory for QCD with two light
flavors is given by \cite{AMY}
\be\label{vis}
\eta \, = \, 86.473 \, \frac{T^3}{g^4 \, \ln (1/g)}
\ee
and is, therefore, {\sl divergent} in the $g \rightarrow 0$
limit. Indeed this indicates importance of non-equilibrium corrections
to Bjorken hydrodynamics, which was pointed out as a possibility
already in the original paper \cite{bj}. The non-equilibrium
corrections become infinitely large in the $g \rightarrow 0$ limit
implying a possible breakdown of Bjorken hydrodynamics.

When viscous corrections are added to Bjorken hydrodynamics, the
energy momentum tensor from \eq{hydrovis1} can be written as (in the
local fluid rest frame) \cite{DGy,Teaney}
\begin{eqnarray}\label{tmnvis}
 T^{\mu\nu} &=& \left( \matrix{ \epsilon (\tau) & 0 & 0 & 0 \cr 0 & p
 (\tau) + \frac{2}{3} \, \frac{\eta}{\tau} & 0 & 0 \cr 0 & 0 & p
 (\tau) + \frac{2}{3} \, \frac{\eta}{\tau} & 0 \cr 0 & 0 & 0 & p
 (\tau) - \frac{4}{3} \, \frac{\eta}{\tau} \cr} \right)\, .
\end{eqnarray}
As follows from \eq{tmnvis}, including viscosity-dependent additive
term in the definition of $T_{\mu\nu}$ reduces the longitudinal
pressure. This may lead to the energy density scaling as shown in
\eq{e1t}, since the pressure $p$ on the right hand side of
\eq{hydroeq} is, in fact, the longitudinal pressure component. In
other words, substituting \eq{tmnvis} into \eq{tcons} modifies
\eq{hydroeq} to give \cite{DGy}
\be\label{hydrovis}
\frac{d \epsilon}{d \tau} \, = \, - \frac{\epsilon + [p - (4/3) \, \eta /\tau ]}{\tau}.
\ee
One can see that shear viscosity tends to reduce the expression in
square brackets on the right hand side of \eq{hydrovis}, making the
scaling of energy density $\epsilon$ with $\tau$ closer to that of
\eq{e1t}. Indeed if shear viscosity is very large this expression in
square brackets may become negative: however, large viscosity needed
for that implies that other non-equilibrium corrections to the
energy-momentum tensor of \eq{hydrovis1} (higher order derivatives of
fluid velocity than the ones included in \eq{hydrovis1}) would also
become important invalidating Eqs. (\ref{tmnvis}) and
(\ref{hydrovis}). When departure from equilibrium becomes this large,
it is not clear whether hydrodynamic approach would remain the best
approximation to the relevant physics. Nevertheless, we have shown
that non-equilibrium corrections tend to bring Bjorken hydrodynamics
closer to the correct $g \rightarrow 0$ limit of the full theory,
indicating that the system at weak coupling is {\sl not} in
equilibrium.

Here we would like to add a word of caution. As was shown in
\cite{hydro}, the most general energy-momentum tensor for a central
collision of two ultrarelativistic nuclei of infinite transverse
extent is given by
\begin{eqnarray}\label{tmngen}
 T^{\mu\nu} &=&  
 \left( \matrix{ \epsilon (\tau) & 0 & 0 & 0 \cr
  0 & p_2 (\tau) & 0 & 0 \cr
  0 & 0 & p_2 (\tau) & 0  \cr
  0 & 0 & 0  & p_3 (\tau) \cr} \right)\, 
\end{eqnarray}
at $z=0$. Therefore the most general $T^{\mu\nu}$ depends only on
three variables: energy density $\epsilon$, and transverse ($p_2$) and
longitudinal ($p_3$) pressure components. The form of the tensors in
Eqs. (\ref{tmnvis}) and (\ref{tmngen}) is very similar: in fact, one
can always express $p_2$ and $p_3$ in terms of $p$ and $\eta$ and vice
versa. Equal number of degrees of freedom in viscous hydrodynamics of
\eq{tmnvis} and in the most general energy-momentum tensor for the
given collision geometry in \eq{tmngen} makes it harder to distinguish
the system really obeying the rules of viscous hydrodynamics from the
system in some non-equilibrium state.

Does our conclusion above indicate that ideal Bjorken hydrodynamics is
impossible to achieve in heavy ion collisions? Possibly yes. However,
one could also imagine the situation where the thermalization and
hydrodynamic behavior of the quark-gluon system are due to
non-perturbative effects, which disappear in the $g \rightarrow 0$
limit. In that case hydrodynamic description would only be correct in
the strong coupling limit of the theory. An example of possible
scaling of energy density could be given by the following ansatz
\be\label{ansatz}
\epsilon \sim \frac{1}{\tau^{1+ \Delta \, \exp (- c /\as)}},
\ee
with $\Delta >0$ and $c > 0$ some non-perturbative constants. Indeed
in the $\as \rightarrow 0$ limit \eq{ansatz} leads to \eq{e1t}. It
would also explain the absence of perturbative $\as \, \ln \tau$
corrections to the scaling of \eq{e1t} observed in \cite{hydro}: the
function in \eq{ansatz} is non-analytic in $\as$ and its Taylor series
has zero coefficients for all non-zero powers of $\as$. The strong
coupling limit of \eq{ansatz}, $\as \rightarrow \infty$, would give
\eq{etau} thus recovering Bjorken hydrodynamics \cite{bj}.

To conclude this Section let us summarize our argument presented here
one more time: the difference of $g \rightarrow 0$ limits of
perturbative QCD and Bjorken hydrodynamics, given by Eqs. (\ref{e1t})
and (\ref{ebt}) correspondingly, indicates that it is impossible for
perturbative evolution of the system to lead to Bjorken
hydrodynamics. Non-perturbative ansatze of the type shown in
\eq{ansatz} are still possible though, indicating that success of hydrodynamic 
description of heavy ion collisions may truly be due to strong coupling
effects, as appears to be demonstrated by RHIC phenomenology
\cite{hydro1,hydro2,Teaney} and as suggested in some thermalization models
\cite{KT05}.

\section{Non-Perturbative Contributions to Jet Quenching and Thermalization}

Below we will try to investigate the role of non-perturbative effects
in heavy ion collisions by modeling them with an instanton
ensemble. Instanton models have been very successful in describing
many features of QCD vacuum. At the same time we are fully aware that
instanton models do not capture {\sl all} the properties of the QCD
vacuum: the best-known example is the inability of instantons to
explain confinement. Therefore, our discussion below describes only
the contribution of instanton fields to the dynamics of heavy ion
collisions, and does not include other non-perturbative (possibly
strong coupling) effects, which may also be very important.

\subsection{Instanton Density in Color Glass Background}

The early stages of a heavy ion collision with proper times $\tau
\lsim 1/Q_s$, are dominated by strong (mostly classical) gluonic
fields produced by the Color Glass Condensates in the colliding nuclei
\cite{mv,k1,jkmw,KMW,KR,GM,KV,K2}. As was shown in \cite{KKL}, these
strong gluon fields modify the vacuum instanton density \cite{size,TE}
\be\label{isize}
n_0 (\rho) \, = \, \frac{0.466 e^{-1.679 N_c}}{(N_c - 1)! (N_c - 2)!} \, 
\frac{1}{\rho^5} \, \left( \frac{2 \pi}{\as (\rho)}
\right)^{2 N_c} \, e^{- \frac{2 \pi}{\as (\rho)}} ,
\ee
to give (see also \cite{CDG})
\be\label{effnf}
n_{sat} (\rho) \, = \, n_0 (\rho) \, \exp \left[ \frac{\pi^3
\rho^4}{\as (N_c^2 - 1)} \, (A|\ G^{a}_{\mu\nu} (x) \,
G^{a}_{\mu\nu} (x) - \, G^{a}_{\mu\nu} (x) \tilde{G}^{a}_{\mu\nu}
(x) \ |A) \right].
\ee
Here $\rho$ is the instanton size, $x$ is the instanton's position in
space-time, $G^{a}_{\mu\nu}$ is the field strength tensor of the
produced gluon field and $(A| \ldots |A)$ denotes averaging over
nuclear wave functions. \eq{effnf} shows that instanton density may be
significantly affected by the strong gluonic fields produced in the
collision.

The dominant gluon field produced in a heavy ion collision in the
saturation/CGC framework is classical \cite{KMW,KR,GM,KV,K2}. Since
the complete classical gluon field is known only numerically
\cite{KV}, following \cite{KKL} we will use the lowest order gluon
field to calculate the correlators in
\eq{effnf}. As was shown in \cite{KKL}, the lowest order ($o (g^3)$) gluon 
field produced in a nuclear collision, which was calculated in
\cite{KR}, gives, in the forward light cone
\ben
\left< G_{\mu\nu}^{a} (x) G_{\mu\nu}^{a} (x) \right>_{LO} \, = \, 
- \frac{g^6}{(2 \pi)^2} \, \frac{A_1 A_2}{S_{1} S_{2}} \,
\frac{C_F}{\pi^3} \, \int \ \frac{d^2 k}{(\underline{k}^2)^2}
\een
\be\label{g24}
\times \, \frac{dk_+ dk_- dk'_+ dk'_-}{(2 \pi)^2} \, \frac{e^{- i k_+
x_{-} - i k_- x_{+} - i k'_+ x_{-} - i k'_- x_{+} } }{(k^2 + i
\epsilon k_0) (k'^2 + i \epsilon k'_0) } \, k \cdot k' ,
\ee
where $A_1$ and $A_2$ are atomic numbers of two nuclei which, for
simplicity, are taken to be cylinders with their axes aligned with the
collision axis and with cross-sectional areas $S_1$ and $S_2$
correspondingly. In arriving at \eq{g24} one has to use the following
formula (see Appendix B of \cite{hydro}, $n, m$ are integers)
\be\label{J7}
\int_{-\infty}^{\infty} d k_+ \, dk_- \, \frac{e^{- i
k_+ x_- - i k_- x_+}}{k^2 + i \epsilon k_0} \, k_+^n \, k_-^m \, = \,
- 2 \, \pi^2 \, \left( \frac{i \, k_T \, \tau}{2 \, x_-} \right)^n \, 
\left( \frac{- i \, k_T \, \tau}{2 \, x_+} \right)^m \, J_{m-n} (k_T \tau)
\ee
to replace each product $k_+ \, k_-$ and $k'_+ \, k'_-$ by $k_T^2 /2$
in anticipation of the integration over $k_+, k_-, k'_+, k'_-$. (If
$n=m$ the power of $(k_+
\, k_-)^n$ on the left hand side of \eq{J7} only brings in a power of 
$(k_T^2 /2)^n$ on its right hand side.) Here $k_T \equiv |{\un k}|$
and in \eq{g24} ${\un k}' = - {\un k}$ \cite{KKL}.

With the help of \eq{J7} we can perform the integration over $k_+,
k_-, k'_+, k'_-$ in \eq{g24} obtaining
\be\label{gg2}
\left< G_{\mu\nu}^{a} (x) G_{\mu\nu}^{a} (x) \right>_{LO} \, = \, 
- 16 \, \as^3 \, C_F \, \frac{A_1 A_2}{S_{1} S_{2}} \, \int \
\frac{d^2 k}{k_T^4} \, k_T^2 \, \left\{ \left[ J_0 (k_T \tau) \right]^2 
- \left[ J_1 (k_T \tau) \right]^2 \right\}.
\ee
Using the fact that, in the same lowest order approximation, the
multiplicity of the produced gluons with transverse momentum $k_T$,
rapidity $y$ and located at impact parameter ${\un b}$ is
\cite{KMW,KR,GM}
\be\label{lomult}
\frac{d N}{d^2 k \, dy \, d^2 b} \, = \, \frac{8 \, \as^3 \, C_F}{\pi} \, 
\frac{A_1 \, A_2}{S_1 \, S_2} \, \frac{1}{k_T^4} \, 
\ln \frac{k_T}{\Lambda},
\ee
and neglecting the logarithm, we rewrite \eq{gg2} approximately as
\be\label{gg3}
\left< G_{\mu\nu}^{a} (x) G_{\mu\nu}^{a} (x) \right>_{LO} \, \approx \, 
- 2 \, \pi \, \int \ d^2 k \, \frac{d N}{d^2 k \, dy \, d^2 b} \,
k_T^2 \, \left\{ \left[ J_0 (k_T \tau) \right]^2 - \left[ J_1 (k_T
\tau) \right]^2 \right\}.
\ee
At the same time, as was shown in \cite{KKL}
\be\label{ggd}
\left< G_{\mu\nu}^{a} (x) \tilde{G}_{\mu\nu}^{a} (x) \right>_{LO} \, = \, 0.
\ee
Combining Eqs. (\ref{gg3}) and (\ref{ggd}) with \eq{effnf} we write
\be\label{effn}
n_{sat} (\rho) \, = \, n_0 (\rho) \, \exp \left\{ - \frac{\pi^3
\rho^4}{\as (N_c^2 - 1)} \, 2 \, \pi \, \int \ d^2 k \, 
\frac{d N}{d^2 k \, dy \, d^2 b} \,
k_T^2 \, \left\{ \left[ J_0 (k_T \tau) \right]^2 - \left[ J_1 (k_T
\tau) \right]^2 \right\} \right\}.
\ee

To analyze \eq{effn} we will use the fact that the gluon transverse
momentum spectrum due to classical fields scales as $\sim 1/k_T^4$ for
$k_T \gsim Q_s$, as shown in \eq{lomult}, and is approximately
constant (up to logarithms) for $k_T \lsim Q_s$
\cite{KMW,KR,GM,KV,K2}. This implies that the $k_T$-integral in \eq{effn} 
is (approximately) dominated by $k_T \approx Q_s$. At early proper
times, for $\tau \lsim 1/Q_s$, the Bessel function $J_1 (k_T \tau)
\approx k_T \tau /2$ can be neglected compared to $J_0 (k_T \tau) \approx 1$ 
in \eq{effn} leading to instanton suppression as was pointed out in
\cite{KKL}. 

On the other hand, at late times, when $\tau \gsim 1/Q_s$, we use the
large-argument asymptotics of Bessel function to write
\be
\left[ J_0 (k_T \tau) \right]^2 - \left[ J_1 (k_T
\tau) \right]^2 \, \approx \, \frac{2}{\pi \, k_T \, \tau} \, \cos \left( 
2 \, k_T \, \tau - \frac{\pi}{2} \right), \hspace*{1cm} k_T \, \tau \gg 1.
\ee
At these late times the exponent of \eq{effn} starts oscillating
around $0$ with a decreasing amplitude. This means that instanton
suppression by produced gluon fields disappears for $\tau \gsim
1/Q_s$.

This disappearance of suppression has a straightforward physical
meaning. As was shown in \cite{hydro}, any (classical and/or quantum)
gluon fields in ultrarelativistic heavy ion collisions lead to the
dominant contribution to the resulting energy density scaling as
\be\label{edens}
\epsilon \, \approx \, \frac{\pi}{2} \,  \int d^2 k \, 
\frac{d N}{d^2 k \, d \eta \, d^2 b} 
\ k_T^2 \, \left\{ \left[ J_1 (k_T \tau) \right]^2 + \left[ J_0 (k_T \tau) \right]^2 
\right\}. 
\ee
At proper times $\tau \gsim 1/Q_s$, \eq{edens} leads to the energy
scaling of \eq{e1t}, corresponding to free streaming
\cite{hydro}. This means that by the proper time $\tau \gsim 1/Q_s$ 
the gluon fields would linearize and propagate as plane waves. Plane
wave gluons have equal chromo-electric and chromo-magnetic fields,
$E^a = B^a$. Remembering that $G_{\mu\nu}^{a \, 2} \, = \, 2 \, ({\vec
B}^{a \, 2} - {\vec E}^{a \, 2})$ we conclude that $G_{\mu\nu}^{a \,
2} \, \approx \, 0$ at times $\tau \gsim 1/Q_s$ and suppression of
instanton density disappears from \eq{effnf}.

\subsection{Instanton Contribution to Jet Quenching}

Let us summarize the emerging picture of the collision. At early
times, $\tau \lsim 1/Q_s$, strong gluon fields are produced and
instantons are suppressed. At later, but still comparatively early
times, $\tau \gsim 1/Q_s$, the gluon fields linearize and vacuum
effects due to instantons are not suppressed anymore. The energy
density of the linearized gluon fields is still quite high at $\tau
\approx 1/Q_s$, since, for $\tau \ge 1/Q_s$
\be\label{ecgc}
\epsilon \, \sim \, \frac{Q_s^3}{\as \, \tau},
\ee
as follows from \eq{edens}. (The factor of $1/\as$ in \eq{ecgc} is due
to the dominance of classical gluon fields at the early stages of the
collision \cite{KMW,KR,GM,KV,K2,KNV,Lappi}.) From here on we will only
perform parametric estimates to demonstrate the principle points:
detailed calculations will be left for future analyses.

We assume that hadronization does not set in until the energy density
of the gluon medium becomes of the order of $\epsilon_h \sim \,
\Lambda^4_{QCD}$. This energy density is obtained by requiring that we 
have one hadron with mass $\sim \Lambda_{QCD}$ per unit volume in the
rest frame of the hadron, which is roughly $\sim
1/\Lambda_{QCD}^3$. Equating the energy density from \eq{ecgc} to
$\Lambda^4_{QCD}$ yields 
\be\label{thad}
\tau_h \, \sim \, \frac{Q_s^3}{\as \, \Lambda_{QCD}^4}.
\ee
\eq{thad} assumes that one-dimensional expansion lasts indefinitely long. 
Of course in a realistic heavy ion collision one dimensional expansion
lasts only up to $\tau \sim R$ with $R$ the nuclear radius. After that
the expansion becomes three-dimensional and the hadronization energy
density $\epsilon_h \sim \, \Lambda^4_{QCD}$ is achieved
faster: we will return to this below.

As we saw in Sect. 3.1, at early times, $\tau \lsim 1/Q_s$, the
non-perturbative instanton effects are suppressed and the vacuum is
perturbative. At later times, $\tau \gsim 1/Q_s$, instantons are not
suppressed anymore and the vacuum is back to being
non-perturbative. Deconfined gluons and quarks propagating through
this vacuum would interact with the gluon (and quark) fields of the
instantons. This may lead to energy loss by these partons in the
instanton ensemble. Quarks and gluons, which, in a purely perturbative
scenario, would have been propagating as plane waves for proper times
$1/Q_s \lsim \tau \le \tau_h$, start interacting with the
non-perturbative vacuum fields loosing their energy. Again this
phenomenon is much more pronounced in $AA$ collisions, since in this
case gluons and quarks spend a much longer time in the deconfined
state interacting with the vacuum fields than they would in $pp$, $pA$
or $e^+ e^-$ collisions. Indeed after hadronization, when the gluons
and quarks would become confined withing hadrons, the energy loss in
vacuum fields would stop: hadrons are eigenstates of the QCD
hamiltonian and they do not loose energy in propagation.

The idea of partonic energy loss in the strong non-perturbative
fields has been previously put forward by Shuryak and Zahed in
\cite{SZ}. However, in their analysis the non-perturbative
fields (sphalerons) were generated in a nuclear collision (see also
\cite{OCS}), and were not the QCD vacuum fields.

To estimate the energy loss in the instanton vacuum fields by a
propagating parton we will employ the results of calculations carried
out in \cite{Bj,BDMPSfull,EL,EL2,Zak,SW,BDMS,GVWZ}. Following
\cite{BDMS} we write the following expression for the energy loss of a 
quark created in a medium and having traveled distance $z$ in that
medium
\be\label{dedz}
- \frac{dE}{dz} \, = \, \frac{\as \, N_c}{2} \, \frac{1}{2 - \alpha}
\, z \, \rho (z) \, \int d^2 q \, q_T^2 \, \frac{d \sigma}{d^2 q},
\ee
where $\rho (z)$ is the density of the medium at point $z$. (Gluon
energy loss is different from \eq{dedz} only by a Casimir.) Here
$\alpha$ is a constant depending on the rate of the medium expansion
with time, $0 < \alpha < 1$ \cite{BDMS}. Since we are interested in
parametric estimates only, the exact value of $\alpha$ is not
important to us. For the same reason the exact form of the
(integrated) scattering cross section on a color charge in the medium
in \eq{dedz} is not important: we will just use the fact that it is
\be
\int d^2 q \, q_T^2 \, \frac{d \sigma}{d^2 q} \, \sim \, \as^2. 
\ee
Using such parametric estimates, we conclude that the integrated
energy loss of a quark or gluon produced in a medium at proper time
$\tau_0$ which travels through the medium up to some hadronization
time $\tau_h$ is
\be\label{de}
- \Delta E \, \sim \, \as^3 \, \int_{\tau_0}^{\tau_h} \, d \tau \, \tau 
\, \rho (\tau).
\ee
Below we will use \eq{de} to estimate the energy loss of a hard parton
in the instanton vacuum fields and in the produced medium.

Of course the energy density of the instanton vacuum is much lower (at
early times) than the energy density of the produced
medium. Therefore, one might assume the energy loss in vacuum fields
to be negligibly small compared to the energy loss in the produced
medium. However, for a wide range of functions $\rho (\tau)$, the
$\tau$-integral in \eq{de} is dominated by {\sl late} proper times
$\tau$ due to an extra factor of $\tau$ in the integrand. Therefore,
for the integrated energy loss, the density of the medium at late
times is more important than the density of the medium at earlier
times. Since, at late times, the density of produced medium becomes
comparable to the instanton density in vacuum, the contributions of
both the medium and the vacuum to partonic energy loss are quite
comparable, as we will see below.

The instanton density in the four-dimensional space-time is
approximately \cite{TE}
\be\label{ninst}
n \, \approx \, 1 \, {\rm fm}^{-4} \, \approx \, \Lambda^4_{QCD}. 
\ee
Assuming that each instanton (or anti-instanton) provides at least one
scattering center for the energy loss calculations of the type carried
out in \cite{Bj,BDMPSfull,EL,EL2,Zak,SW,BDMS,GVWZ}, \eq{ninst} allows
us to put a lower bound on the resulting density of scattering centers
at
\be\label{rinst}
\rho_{inst} (\tau) \, \approx \, \Lambda^3_{QCD}.
\ee
Substituting \eq{rinst} into \eq{de} yields the following energy loss
in the instanton vacuum fields ($\tau_h \gg \tau_0$)
\be\label{instel}
- \Delta E_{inst} \, \sim \, \as^3 \, \Lambda^3_{QCD} \, \tau_h^2.
\ee

To see whether the energy loss given by our estimates in \eq{instel}
is large or small we want to compare it to the energy loss in the
quark-gluon plasma. Indeed, as we argued above, this thermalized
medium (QGP) can not be generated perturbatively. Below we will
investigate whether parton-instanton interactions can lead to creation
of a thermalized medium at early times. In the meantime we would just
assume that a thermal medium with the energy density scaling given by
\eq{etauid} has been created in the collision. For this medium,
requiring that the energy density should map onto \eq{ecgc} at $\tau =
1/Q_s$, we write
\be\label{e1}
\epsilon_{qgp} \, \sim \, \frac{Q_s^4}{\as \, (Q_s \, \tau)^{4/3}}.
\ee
Remembering that if $\epsilon \sim T^4$ then $\rho \sim T^3$ we deduce
from \eq{e1} the density of the produced medium
\be\label{r1}
\rho_{qgp} (\tau) \, \sim \, \frac{Q_s^3}{\as^{3/4} \, Q_s \, \tau}.
\ee
Substituting \eq{r1} into \eq{de} gives us a parametric estimate of
energy loss for a parton in quark-gluon plasma
\be\label{qgpel}
- \Delta E_{qgp} \, \sim \, \as^{9/4} \, Q_s^2 \, \tau_h.
\ee

Now we can compare the energy loss in the instanton vacuum from
\eq{instel} to the energy loss in QGP from \eq{qgpel}. First, assuming 
that nuclei have infinite transverse extent, such that the
hadronization time is given by \eq{thad}, we substitute it into
Eqs. (\ref{instel}) and (\ref{qgpel}) to obtain
\be\label{instel1}
- \Delta E_{inst} \, \sim \, \as \, \frac{Q_s^6}{\Lambda_{QCD}^5} 
\ee
and
\be\label{qgpel1}
- \Delta E_{qgp} \, \sim \, \as^{5/4} \, \frac{Q_s^5}{\Lambda_{QCD}^4}. 
\ee
Since $Q_s \gg \Lambda_{QCD}$ and $\as \ll 1$ we conclude that
\be\label{ivsqgp}
|\Delta E_{inst}| \, \gg \, |\Delta E_{qgp}|,
\ee
i.e., that the energy loss in the instanton vacuum is actually
somewhat greater than the energy loss in the produced medium! Since
the saturation scale grows with energy and atomic number, $Q_s^2 \,
\sim \, A^{1/3} \, s^{\lambda /2}$, the instanton-induced energy loss 
would increase with energy and the system's size faster than the
energy loss in a medium.

However, the conclusion of \eq{ivsqgp} was reached for the
one-dimensional expansion only. For realistic nuclei one-dimensional
expansion stops at $\tau \approx R$. Since the hadronization time
$\tau_h$ from \eq{thad} is parametrically much larger than $R$, the
one-dimensional expansion stops before the time $\tau_h$ is
achieved. After that the expansion becomes three-dimensional. Assuming
that free streaming continues in the 3D expansion, the energy density
for the proper times after $\tau \approx R$ is given by
\be\label{e3d1}
\epsilon \, = \, \epsilon (\tau = R) \, \left( \frac{R}{\tau} \right)^3
\ee
where $\epsilon (\tau = R)$ could be obtained by putting $\tau = R$ in
\eq{ecgc}. \eq{e3d1} becomes
\be\label{e3d2}
\epsilon \, \sim \, \frac{Q_s^3}{\as \, R} \, \left( \frac{R}{\tau} \right)^3.
\ee
Equating the energy density from \eq{e3d2} to $\epsilon_h \sim
\Lambda^4_{QCD}$ we obtain the hadronization time for a realistic nuclear 
collision
\be\label{thadr}
\tau_h^{3D} \, \sim \, R \, \frac{Q_s}{\Lambda_{QCD}} \, 
\frac{1}{(\as \, R \, \Lambda_{QCD})^{1/3}}.
\ee
If we use the hadronization time from \eq{thadr} in
Eqs. (\ref{instel}) and (\ref{qgpel}) we get
\be\label{instel2}
- \Delta E_{inst} \, \sim \, \as^{7/3} \, \frac{Q_s^2}{\Lambda_{QCD}}
\, (R \, \Lambda_{QCD})^{4/3}
\ee
and
\be\label{qgpel2}
- \Delta E_{qgp} \, \sim \, \as^{23/12} \, \frac{Q_s^3}{\Lambda_{QCD}^2}
\, (R \, \Lambda_{QCD})^{2/3}.
\ee
To compare Eqs. (\ref{instel2}) and (\ref{qgpel2}) we first analyze
them in the quasi-classical approximation: this regime seems to be
adequate for mid-rapidity $AA$ collisions at RHIC. As in the
quasi-classical McLerran-Venugopalan model $Q_s^2 \approx A^{1/3}
\Lambda^2_{QCD}$ and since $R \, \Lambda_{QCD} \sim A^{1/3}$, we can
conclude, neglecting the factors of $\as$ in the front, that for
Eqs. (\ref{instel2}) and (\ref{qgpel2})
\be\label{mv3d}
|\Delta E_{inst}| \, \sim A^{7/9} \, > \, |\Delta E_{qgp}| \, \sim \,
A^{13/18}.
\ee
Again, the instanton-induced energy loss appears to be slightly higher
than the one in the plasma. Indeed, when quantum evolution corrections
are included the saturation scale gets extra enhancement as $Q_s^2
\sim s^{\lambda/2}$: in that case $|\Delta E_{qgp}|$ from \eq{qgpel2}
would grow faster with energy than $|\Delta E_{inst}|$ from
\eq{instel2}, eventually becoming larger. Still we have shown that the 
energy loss in the instanton vacuum is parametrically comparable to
the energy loss in QGP even for realistic nuclear collisions, except
for very high energies when $Q_s \gg \Lambda_{QCD} \, (R \,
\Lambda_{QCD})^{2/3}$. The instanton energy loss from both
Eqs. (\ref{instel1}) and (\ref{instel2}) also increases with energy
since $Q_s^2 \sim s^{\lambda/2}$, which is in qualitative agreement
with the absence of suppression of hadron production at lower (SPS)
energy $AA$ collisions and with the onset of suppression at higher
(RHIC) energies.

Indeed there is no real physical competition between the energy loss
in QGP and in the instanton vacuum. If QGP is formed, instantons would
be suppressed \cite{GPY}, much like suppression in CGC that was
discussed in Sect. 3.1. Here we calculate energy loss in QGP just to
have a reference to compare instanton energy loss to. We need to do
this since all our estimates in this Section are parametric and can
not be used to produce numbers to compare to experimental data.

Let us also note that a significant energy loss due to instanton
fields only takes place in $AA$ collisions. For instance, in $pA$
collisions energy loss is given by Eqs. (\ref{instel2}) and
(\ref{qgpel2}) with the nuclear radius $R$ replaced by the proton
radius, $\tau_h^p \sim r_p \sim 1/\Lambda_{QCD}$ and the nuclear
saturation scale $Q_s$ replaced by the proton saturation scale, which,
in the quasi-classical approximation is roughly $\sim
\Lambda_{QCD}$. After performing these substitutions we get
\be
- \Delta E^{pA}_{inst} \, \sim \, \Lambda_{QCD} \, \sim \, - \Delta
E^{pA}_{qgp},
\ee
which is very small, much smaller than the energy loss in $AA$
collisions given by Eqs. (\ref{instel2}) and (\ref{qgpel2}).

To summarize, we have succeeded in showing that the energy loss in the
vacuum fields modeled by an instanton ensemble is appreciably high,
being parametrically comparable to the energy loss expected in hot
quark-gluon plasma. Once again, let us reiterate that such conclusion
is only possible due to a peculiar form of the integrated energy loss
in \eq{de}, which was derived in \cite{BDMS,GVWZ}, making $\Delta E$
mostly sensitive to the color charge density at late proper times. The
obtained instanton energy loss increases with both center of mass
energy of the collisions and with the size of the system.

\subsection{Instanton Contribution to Thermalization}

Finally, let us address the following question: if the interactions of
the quarks and gluons with the instanton vacuum fields are
sufficiently strong to generate a significant energy loss, as shown in
Eqs. (\ref{instel1}) and (\ref{instel2}), should not the same
interactions lead to thermalization of the quark-gluon medium? This is
a difficult question which, in general, requires understanding to what
extent interactions of partons with the instanton fields would trigger
interactions of partons with each other, which, in a purely
perturbative approach \cite{KV,hydro}, would stop by $\tau \approx
1/Q_s$. Here we will postpone this important question for further
investigation, which is beyond the scope of this work, and concentrate
on parton-instanton interactions only.

As was shown in \cite{KV,hydro} the perturbative gluon fields are
produced by $\tau \approx 1/Q_s$ with zero longitudinal pressure
component. Namely, at $z=0$ and at $\tau \approx 1/Q_s$ the
energy-momentum tensor looks (approximately) like $T_{\mu\nu} =
\mbox{diag} (\epsilon, p , p, 0)$. Thermalization of the system must 
be accompanied by generation of the non-zero longitudinal pressure
component comparable to the transverse pressure components
\cite{ALMY}. This is a necessary, though not sufficient, condition 
for thermalization. Let us estimate when such a longitudinal pressure
component would be developed in parton-instanton interactions: that
would give us a lower bound on thermalization time.

Avoiding all the subtleties of parton-instanton interactions
\cite{heinst} we assume that the upper bound on the cross section of a 
parton-instanton interaction is given by the instanton size squared,
$\sigma_{inst} \, \sim \, \langle \rho \rangle^2 \, \sim \,
1/\Lambda^2_{QCD}$ \cite{TE}. The mean free path of a parton in the
instanton medium with the density given by \eq{rinst} is then
\be
\lambda_{inst} \, = \, \frac{1}{\rho_{inst} \ \sigma_{inst}} \, \sim \, 1/\Lambda_{QCD}. 
\ee
In each parton-instanton interaction, the momentum transfer to the
parton would be of the order or $1/\langle \rho \rangle \sim
\Lambda_{QCD}$. After $N$ rescatterings the parton's longitudinal momentum 
gets broadened to approximately
\be
p_z \, \sim \, \sqrt{N} \, \Lambda_{QCD}.
\ee
For the system to become isotropic we demand that $p_z \sim p_T \sim
Q_s$, which gives $N \sim Q_s^2 /\Lambda^2_{QCD}$. The time it takes
to have $N \sim Q_s^2 /\Lambda^2_{QCD}$ parton-instanton rescatterings
is roughly given by
\be\label{tth}
\tau_{th} \, \sim \, \lambda_{inst} \, N \, \sim \, \frac{Q_s^2}{\Lambda^3_{QCD}}.
\ee
This is a very late thermalization time. Even in McLerran-Venugopalan
model, where $Q_s^2 \approx A^{1/3} \Lambda^2_{QCD}$, this
thermalization time becomes comparable to the nuclear radius,
$\tau_{th} \sim R$, which corresponds to the end of the
one-dimensional expansion. Certainly such thermalization time is too
late for hydrodynamic models describing RHIC data
\cite{hydro1,hydro2}. Including energy dependence in $Q_s$ would only
make $\tau_{th}$ larger, leading to thermalization time which
increases with energy. Indeed a numerical prefactor in \eq{tth}, when
calculated, may change the actual numerical results
somewhat. Nevertheless we have demonstrated that, at least for very
large nuclei, instanton vacuum can significantly contribute to jet
quenching, as was shown in \eq{instel2}, but is not likely to bring in
fast thermalization.

\section{Conclusions}

Above we have presented two arguments demonstrating that perturbative
thermalization leading to ideal Bjorken hydrodynamics is impossible in
heavy ion collisions. The only possible way to reconcile this with the
phenomenological evidence for fast equilibration of the quark-gluon
system produced in the collision \cite{hydro1,hydro2} is to suggest
that thermalization takes place through some non-perturbative
processes. It might be reasonable to assume that non-perturbative
effects become important at proper times $\tau_\Lambda \, \approx \,
1/\Lambda_{QCD}$ in $AA$ collisions, as they are known to do in $pp$,
$pA$ and $e^+ e^-$ collisions. Non-perturbative effects involve QCD
strings \cite{strings} with the string tension $\sigma \approx
1$~GeV/fm. Gluons and quarks produced in the initial stages of the
collision carry a typical transverse momentum of the order of $Q_s$,
which is $Q_s = 1 \div 1.4$~GeV at RHIC \cite{KN}. This number is
comparable to $\sigma \, \tau_\Lambda \, \approx \, 1$~GeV, indicating
that QCD string effects may become important at proper time
$\tau_\Lambda$, introducing strong-coupling interactions between the
partons and, possibly, bringing the system towards fast
thermalization. The corresponding time scale $\tau_\Lambda \, \approx
\, 1/\Lambda_{QCD} \, \approx \, 1$~fm/c appears to be consistent with the thermalization
time of approximately $0.5$~fm/c required by hydrodynamical models
\cite{hydro1,hydro2} to describe RHIC elliptic flow data \cite{flow}.

Non-perturbative thermalization, either through instantons or through
QCD string interactions, would not violate our argument of
impossibility of perturbative thermalization presented above and in
\cite{hydro}. As we have argued in Sect. 2.2, thermalization in heavy 
ion collisions could only be a strong coupling phenomenon: such
effects lie outside of the analysis carried out in \cite{hydro}. The
small coupling $\as$ used in the energy loss discussion of Sect. 3.2
is due to parton momentum broadening in the non-perturbative instanton
vacuum and does not infringe on the non-perturbative nature of the
vacuum. Indeed, the thermalization scenario discussed in Sect. 3.3 is
essentially partonic, but it requires non-perturbative background
instanton fields, which are also outside of the perturbative analysis
of \cite{hydro}.

What appears to be unclear in non-perturbative thermalization
scenarios with either instanton or string interactions is the energy
dependence of thermalization time. It is not clear how string-induced
interactions "know" about the center-of-mass energy of the
collision. Instanton-mediated thermalization appears to have
thermalization time (\ref{tth}) increasing with energy. This
contradicts the fact that thermalization time required by
phenomenology at RHIC is short, while it is not clear whether
thermalization has ever occurred at the SPS implying a much longer
thermalization time at lower energies.

Non-perturbative effects appear to be needed to describe jet quenching
in heavy ion collisions as well. As we discussed above in Sect. 3.1,
in the purely perturbative scenario of the collision, by the proper
time $\tau \sim 1/Q_s$ the gluon fields would linearize and propagate
as plane waves. (Similar results have been obtained for quark fields.) 
This result implies that, as long as perturbative interactions are
concerned, nothing happens to the gluons after $\tau
\sim 1/Q_s$. Therefore, the resulting gluon spectra would be
completely determined by the (very) early stage dynamics. This would
be in contradiction to the observation of Cronin enhancement at
mid-rapidity in $d+Au$ collisions at RHIC \cite{rhic_dA,brahms-2} and
the observation of suppression of hadron production at mid-rapidity in
$Au+Au$ collisions \cite{rhic_aa}: all the known perturbative
production mechanisms imply that, since hadrons produced at
mid-rapidity in $d+Au$ collisions exhibit Cronin enhancement, so
should the partons in the early stages of $Au+Au$
collisions. Therefore, to reconcile RHIC data with the theory one has
to assume that strong final state interactions take place in $AA$
collisions, which could only be non-perturbative as was demonstrated
by the above analysis. Such effects could be due to non-perturbative
thermalization leading to formation of QGP in which the partons would
loose energy. Alternatively, as we have shown above, partons can loose
energy in the non-perturbative vacuum fields, to which they are
exposed until the hadronization time in heavy ion collisions.

To summarize we note that we have demonstrated above that interactions
in the final stages of heavy ion collisions can only be
non-perturbative in nature. They may or may not lead to rapid
thermalization, but they are likely to significantly contribute to jet
quenching.

\section*{Acknowledgments}

The author would like to thank Peter Arnold, Eric Braaten, and Ulrich
Heinz for many informative discussions.

This work is supported in part by the U.S. Department of Energy under
Grant No. DE-FG02-05ER41377.

\end{document}